# Symbol Dynamics, Information theory and Complexity of Economic time series

Geoffrey Ducournau[1]

## Abstract

We propose to examine the predictability and the complexity characteristics of the Standard&Poor500's dynamics behaviors in a coarse-grained way using the symbolic dynamics method and under the prism of the Information theory through the concept of entropy and uncertainty. We believe that experimental measurement of entropy as a way of examining the complexity of a system is more relevant than more common tests of universality in the transition to chaos because it does not make any prior prejudices on the underlying causes associated with the system's dynamics, whether deterministic or stochastic. We regard the studied economic time series as being complex and propose to express it in terms of the amount of information this last is producing on different time scales and according to various scaling parameters.

## 1. Introduction

Over the last decades, advances in the scientific study of chaos in the understanding of the behavior of dynamic systems have been an important motivator of the modern study of complex systems such as economic time series. Until now, one of the most useful measures of complexity has been the Lyapunov exponent which measures the average rate of divergence or convergence of two nearby trajectories. For example, given time series data obtained from unknown economic time series, the Lyapunov exponent may be estimated using the phase space reconstruction method [14] or by a method based on nonparametric regression [4], a positive Lyapunov exponent is generally taken as evidence of the chaotic character of the underlying system, and consequently, conclude on the deterministic specificity of the system.

However, the underlying problem with the previous method relies on the fact that it assimilates chaos to complexity without making any distinction and therefore addresses measures of complexity that are suitable for deterministic dynamic systems to perhaps non-deterministic systems.

Indeed, it is obvious that it exists confusion about the relationship between chaos and complexity. Chaos deals with deterministic systems whose trajectories diverge exponentially over time, and it is well known that this property can be found in the behavior of complex systems. However, Complex Systems do not necessarily have a natural representation of this form, or these behaviors, and can be simply described as a system with many degrees of freedom that are partially but not completely independent. Moreover, it is known that the non-linearity of complex systems are the consequences of interactions with their respective environment, most of them are what we called open and dissipative system, and it is not entirely clear that system-environment interactions are modeled by chaos since the deterministic equations of chaos can be most easily thought as describing a closed system [19].

[1]Geoffrey Ducournau, PhDs, Institute of Economics, University of Montpellier; G.ducournau.voisin@gmail.com

Thus, contrary to chaos, complex systems are dealing with both the structure of the dynamics of the systems and their interactions with their environment. We can also add that chaos is commonly referred to the notion of disorder and randomness, as complex systems that are often described as physical phenomenon that are situated between two extremes, the Order and the Disorder. However, stochastic systems can perfectly emphasize such characteristics although they are definitely not governed by deterministic equations. Indeed, if we take as an example the methodology used by Shannon in 1943 [2] to illuminate the structure of a message, we understand that languages such as English or Chinese are complex systems resulting in a series of symbols (events) that can be turned to a stochastic process that is neither deterministic (the next symbol cannot be calculated with certainty) nor random (the next symbol is not totally free). Languages are governed by a set of probabilities; each symbol has a probability that depends on the state of the system and also perhaps on its previous history.

Consequently, it is crucial to avoid cofounding chaos with complex systems since they are intrinsically distinct. If chaos implies complexity, complex systems are not necessarily governed by chaotic systems. However, defining what is a complex system remains a challenge, we can so far only characterize a complex system by some properties [10] such as : a complex system consists of many components, these components can be either regular or irregular, these components exits on different scales (spatial and time scales) leading the system to exhibit mono-fractality or multifractality structures [3], the system is locally auto-adaptative, exhibits self-organization and hierarchy of structures [5].

The purpose of this paper is to propose another approach in measuring and quantifying the degree of memory length and complexity of a system such as an economic time series, without the premise of considering it as deterministic if complexity there is. To this end, we propose a similar method proposed by Shannon relying on information theory consisting in considering the degree of complexity of a message as a measure of uncertainty. Roughly speaking, Shannon considered that the more order exists in a sample of English text (order in a form of statistical patterns), the more predictability there is, and in Shannon's terms, the less information is conveyed by each subsequent letter.
We propose to use the same approach and applies it to the economics time series Standard&Poor500, to study the degree of uncertainty the dynamics of prices changes displayed on different time scales (minutes, 4-hours and daily) and to measure the degree of predictability of changes in log returns according to different time scales.

To that end, we organize this paper into different sections. Section 2 will introduce the Symbol dynamics studies which is a key element of information theory. In section 3, we show evidence of the character non-Markovian of the Standard&Poor500 on certain time scales and conclude on the presence of memory process and on the degree of predictability of this last. In section 4 we propose to quantify the degree of complexity of Standard&Poor500 price dynamics through the measure of the Generalized Rényi dimension and entropy, and also by highlighting its singularity spectrum. Finally, we conclude this paper by highlighting keys results from this paper.

## 2. Symbolic dynamics and partitions: methodology

Symbolic dynamics study complex system based on the symbol sequences obtained for a suitable partition of the state space. The basic idea behind symbolic dynamics is to divide the phase space of a system into a finite number of regions (partitions) and to label each region by a symbol whether alphabetical or numerical [16].

In this paper we investigate the economic time series of the American Index Standard&Poor500, a weighted measurement stock market of the 500 largest companies, more specifically, we will consider the log-returns of this index price as the studied system and we call $r_i = \log(P_{i+1}) - \log(P_i)$ for $i = 0, 1, \ldots, N-1$, $r_i$ the returns for every discrete time $i$ and $P_i$ the index price for respective discrete time $i$. The dataset starts in January 2016 and ends in March 2021. Moreover, different time scales will be studied and analyzed especially small-time scales (minutes), medium time scales (4-hours), and larger time scales (daily). The purpose is to emphasize the difference of behaviors and complexity according to different time scales. Indeed, as mentioned above, if the economic time series Standard&Poor500 is a complex system, then it should characterize itself by exhibiting local structures (patterns, statistics) on smaller time and space scales that are self-affine or self-similar to structure on larger time and space scales.

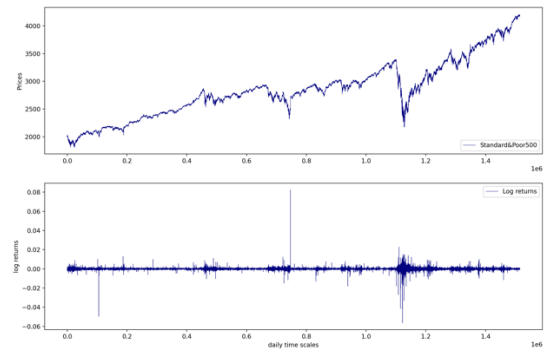

**Figure1:** Standard&Poor500 prices and log returns from January 2016 to March 2021 on minute time scales.

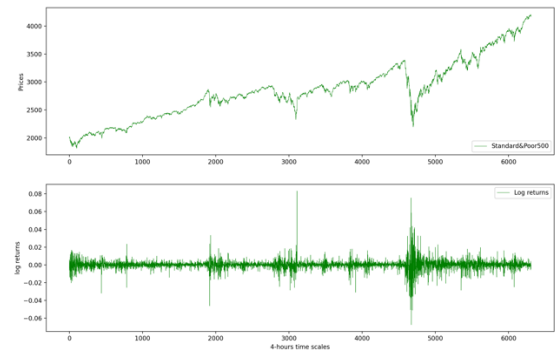

**Figure2**: Standard&Poor500 prices and log returns from January 2016 to March 2021 on 4-hours time scales.

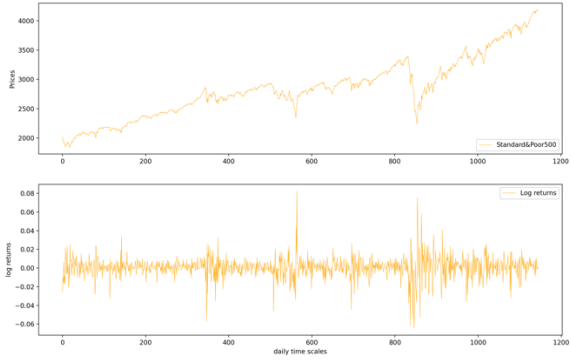

**Figure3**: Standard&Poor500 prices and log returns from January 2016 to March 2021 on daily time scales.

First, the method of symbolic dynamics consists of generating a symbol sequence by a proper partition of the phase space. In our case, the symbolic dynamics will come from a number of binary choices of the state space namely $\{0, 1\}$.

In other words, if we consider the phase space interval $(-\infty, +\infty)$, for every given discrete-time $i$, when $r_i \leq 0$ then $r_i = 0$ and when $r_i \geq 0$, then $r_i = 1$. Consequently, we can define a new vector called $x_i$ and we allocate to each $x_i$ for $i = 0, 1, \ldots, N-1$ an attributed symbol sequence such that: $x_i = x_0, x_1, \ldots, x_{N-1}$ with $x_i \in \{0,1\}$.

Second, we naturally want to divide the state space of the symbol sequence $x_i$ into different partitions or segments that we call $S$ of equal length $L$, in which the dynamics of the complex system takes place in order to measure the possible time-dependent information in the partition. To this end we will use a sliding window of given length $L$ moving from one step by one step from $x_0$ to $x_{N-1}$.

Then we define $S_j^{(L)} = \{S_0^{(L)}, S_1^{(L)}, \ldots, S_R^{(L)}\}$ with $S_j^{(L)}$ corresponding to the segment $j$ with respective size $L$ and $j = 0, 1, \ldots, R$ with $R$ correspond to the total number of segments $S_j$ for a certain length $L$. For example, if we choose $L = 3$, then we have:

$$S_j^{(L=3)} = \{S_0^{(3)}, S_1^{(3)}, \ldots, S_R^{(3)}\} \text{ with :}$$

$$S_0^{(3)} = \{x_0, x_1, x_2\}$$
$$S_1^{(3)} = \{x_2, x_3, x_4\},$$
$$\ldots$$
$$S_R^{(3)} = \{x_{N-3}, x_{N-2}, x_{N-1}\}.$$

For a question of readability, we will define:

$$S_j^{(L)} = \{x_{j,l}\}^{(L)} \quad (2.1),$$

with $j = 0, \ldots, R$ the $j$-ème partition and $l = 1, 2, \ldots, L$ with $L$ the size of each partition. Thus, for $L = 3$ we have:

$$S_j^{(3)} = \{x_{0,1}, x_{0,2}, x_{0,3}\}, \ldots, \{x_{R,1}, x_{R,2}, x_{R,3}\}.$$

Because each symbol $x_{j,l}$ can take only two different values 0 or 1, therefore the number of possible partitions $S$ for any given $L$ is defined by the following relation:

$$\Phi(L) = 2^L \quad (2.2).$$

Regarding the previous example, if $L = 3$ then $\Phi(L = 3) = 8$ and:

$$S^{(L=3)} = [\{0\ 0\ 0\}, \{0\ 0\ 1\}, \{0\ 1\ 1\}, \{1\ 1\ 1\},$$
$$\{1\ 1\ 0\}, \{1\ 0\ 0\}, \{1\ 0\ 1\}, \{0\ 1\ 0\}].$$

Third, now that we have divided the symbol sequence into a set of partitions of equal length constituting of 0 and 1, we understand that we have different states $\Phi(L)$ possible, states that we also called patterns. We are not interested in these ordinal patterns themselves but rather in their frequency distribution, namely, how often are we likely to meet these ordinal patterns. In this fashion, the main point is to summarize the original economic time series into a set of relative frequencies, and then use this set for conducting a study based on information theory. Indeed, we can easily acquire the probabilities of each partition by computing the frequencies of different symbol sequences that occur as follows:

$$p_j^{(L)} = \frac{\phi(S^{(L)})}{R} \quad (2.3),$$

with $j = 1, \ldots, R$, $\phi(S^{(L)})$ is the number of times a segment $S$ of size $L$ appears within the economic time series. In the same way, if we want to know the probability of every possible segments independently of the one coming from the economic time scales itself, in that case we will define:

$$p_k^{(L)} = \frac{\phi^L}{R} \quad (2.4),$$

with $k = 1, \ldots, \Phi(L)$ where $k$ is the number of all possible sequence of size $L$ and $\phi$ is the number of time a certain sequence $x_0, x_1, \ldots, x_{N-1}$ appears.

Fourth, now that we have a symbol series divided into a set of partitions, and that we understand how to quantify the degree of information from a partition based on the information theory in a sense that each partition is seen as an event with given probabilities, we want to be able to analyze these different probabilities, which is not easy to compute when we deal with sequences of series of binaries numbers. To go over this issue, we propose to convert these series of binary numbers into fractional numbers. Let's define $\pi$ a fractional number associated with every given partition $S^{(L)}$ such as:

$$\pi\left(S_j^{(L)}\right) = \sum_{l=1}^{L} x_{j,l}\, 2^{-l} \quad (2.5),$$

for $j = 0, 1, \ldots, R$ and $0 \leq \pi\left(S_j^{(L)}\right) < 1$. Once again, if we take into account the above example with $L = 3$, we obtain the new sequence of symbols such that:

$$\pi(S^{(L=3)}) = \{0, 0.125,\ 0.375, 0.875, 0.75,\ 0.5,\ 0.625, 0.25\}.$$

We are now able to substitute each symbol sequence (partition) by a fractional and unique number between 0 and 1, to which we assigned a given probability that corresponds to its frequency of occurrence within the economic time series. The probability defined by equation (2.3) now become:

$$p_j^{(L)} = \frac{\phi\left(\pi(S^{(L)})\right)}{R} \quad (2.6),$$

with $j = 1, \ldots, R$, $\phi\left(\pi(S^{(L)})\right)$ is the number of times a fractional number $\pi(S^{(L)})$ from a partition of size $L$ appears within the economic time series.

This will be the subject of section 3.

## 3. Evidence of memory and predictability

One method to conclude about the presence of a long memory process within a sequence of symbols is to investigate the character Markovian or non-Markovian of the series. The Markov property states that the conditional probability distribution for the studied system at the next step depends only on the current state of the system and not additionally on the state of the system at previous steps. By considering the series of sequences of fractional numbers defined in section 2 by equation (2.5) if the sequence $\pi\left(S_j^{(L)}\right)$ is defined by Markov property, thus we should have:

$$P\left(\pi\left(S^{(L)}_{j+1}\right)\Big|\pi\left(S^{(L)}_{j}\right),\pi\left(S^{(L)}_{j-1}\right),\ldots,\pi\left(S^{(L)}_{0}\right)\right)$$

$$= P\left(\pi\left(S^{(L)}_{j+1}\right)\Big|\pi\left(S^{(L)}_{j}\right)\right).$$

Before exhibiting the conditional probabilities of each fractional number $\pi\left(S_j^{(L)}\right)$ given the collections of subsets, we propose to compare the probabilities of the empirical symbol sequences with perfectly random sequences for given partition length $L = \{2, 4, 6, 8\}$, and test their stationarity. The purpose is to give evidence that conditional probabilities are a more reliable method than stationary tests when we want to detect the memory process within the symbol sequence series.

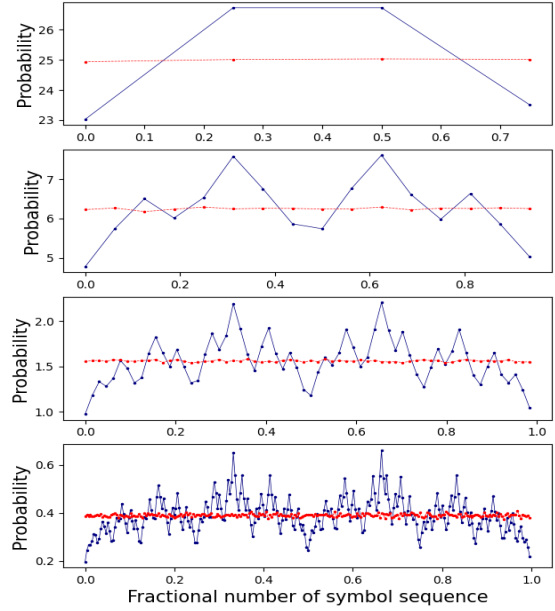

**Figure4:** Probabilities of the minutes log returns dynamics for partition of length $L = [2, 4, 6, 8]$. Therefore, each series have respectively 4, 16, 64 and 256 possible type of symbols. In blue we represent the dynamics symbols sequences of the Standard&Poor500, in red we represent a random dynamics symbols sequence.

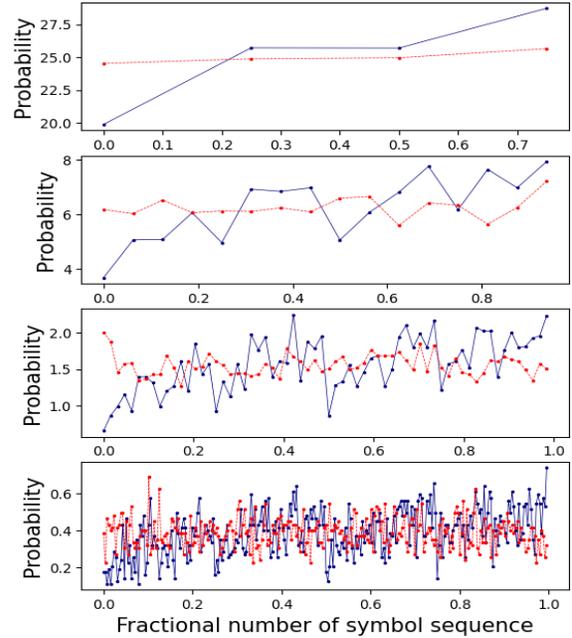

**Figure5:** Probabilities of the 4-hours time scales log returns dynamics for partition of length $L = [2, 4, 6, 8]$. Therefore, each series have respectively 4, 16, 64 and 256 possible type of symbols. In blue we represent the dynamics symbols sequences of the Standard&Poor500, in red we represent a random dynamics symbols sequence.

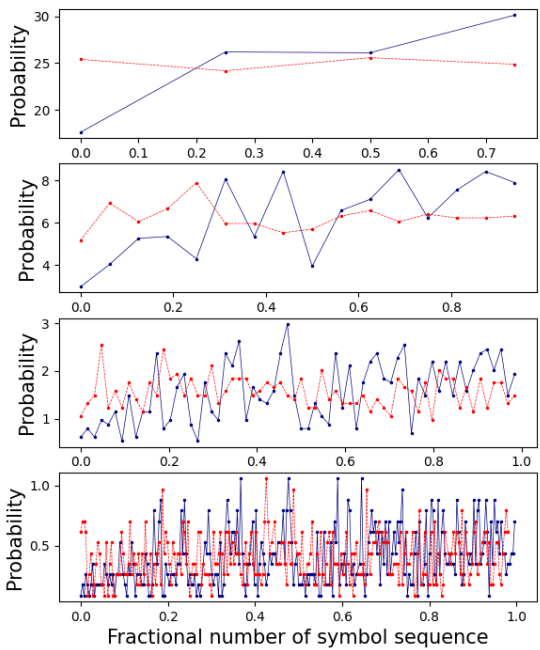

**Figure6:** Probabilities of the daily scales log returns dynamics for partition of length $L = [2, 4, 6, 8]$. Therefore, each series have respectively 4, 16, 64 and 256 possible type of symbols. In blue we represent the dynamics symbols sequences of the Standard&Poor500, in red we represent a random dynamics symbols sequence.

Figure4,5,6 emphasized the probabilities of log returns dynamics for symbol sequence for different length $L$ of the Standard&Poor500 against random dynamics symbol sequence of the same length, on different respective time scales namely minutes, 4-hours, and daily. The differences between the log-returns symbol sequence and the random symbol sequence on each time scales are clearly visible. On minutes time scales we assess the emergence of clear patterns and hierarchical structures that are redundant. On larger time scales (i.e. 4-hours and daily), the probabilities of symbol sequences of the Standard&Poor500 are still different from random symbol sequences, but we also observe less emergence of ordered structures and patterns. Moreover, when the length $L$ is increasing, symbol sequences from larger time scales exhibit behaviors less distinguishable from a purely random process. Thus, our first deduction is that the underlying Stadard&Poor500 dynamics are exhibiting memory length.

We perform an Augmented-Dickey-Fuller-test to test the stationary of symbol sequences and the results are highlighted within Table1.

| ADF test | Critical Values | L=2 | L=4 | L=6 | L=8 |
|---|---|---|---|---|---|
| Minute | ADF | -1.90 | -9.67 | -1.86 | -3.56 |
| 4hours | ADF | -1.65 | -0.49 | -1.82 | -1.82 |
| Daily | ADF | -1.89 | -3.36 | -1.66 | -8.28 |
|  | 1% | -10.4 | -4.47 | -3.55 | -3.46 |
|  | 5% | -5.78 | -3.28 | -2.91 | -2.87 |
|  | 10% | -3.39 | -2.77 | -2.60 | -2.57 |

**Table1:** Augmented-Dickey-Fuller test as stationary test run on Standard&Poor500 dynamics symbol sequences for $L = [2, 4, 6, 8]$.

From Table1 we assess that for any given length $L$, on every time scales the dynamics of symbols sequences are stationary with negative ADF values.

The fact this stationary test rejects the null hypothesis cannot however enable us to give any conclusion on the presence or not of the memory process within the series.

First, because stationary test makes the assumption [17] that the sample to be tested spans over a time range that is longer than the memory of the underlying process. Consequently, when the stationary test typically fails, one of the multiple assumptions could be to blame: the memory of the underlying process could be much longer than the sample size, or in other words, the sample could be too short to characterize the underlying process. Thus, the only way to conclude non-stationarity from a unit-root test failure is to treat the other hypotheses as axioms, meaning that not only the non-stationarity is considering as sole null hypothesis (which is the case for stationary test), but by considering the null hypothesis as a combination of hypotheses, i.e., the non-stationarity, a specific diffusion model and the memory of the underlying process is not longer than the range of the sample tested.

Second, if memory and predictability are tightly related, it is hard to conclude on memory based on stationarity or non-stationarity. Even though advances academic book of econometrics [11] claims that stationary series are memoryless, it is not always necessarily the case. And actually, a timeseries can both be stationary and exhibit a lot of memory, as is the case of stationary Gaussian processes [9, 6], as a timeseries can be both memoryless and non-stationary [11].

Therefore, we prefer to deal with the problem of memory by the following qualitative question: does knowing the past reduce our uncertainty about the future?

Thus, we are more interested in conditional probabilities to analyze the symbol sequence as defined above by this relation:

$$P\left(\pi\left(S^{(L)}_{j+1}\right) \Big| \pi\left(S^{(L)}_{j}\right), \pi\left(S^{(L)}_{j-1}\right), \ldots, \pi(S^{(L)}_{0})\right).$$

We call $\Omega$ the sample space such that $\Omega = \left\{\left(S^{(L)}_{j}\right), \pi\left(S^{(L)}_{j-1}\right), \ldots, \pi(S^{(L)}_{0})\right\}$, $\mathcal{F}$ the filtration that represent a set of events on $\Omega$ and $P$ the probability of each event. Then, the conditional probabilities of the symbol sequence must satisfy the following constraint:

$$\sum_{S^{(L)} \in \Omega} P\left(S^{(L)} \mid \mathcal{F}\right) = 1 \quad (2.7).$$

Conditional probability is a crucial indicator in the discussion of the Markovian property. From the definition of the discrete-time Markov chains, one has:

$$\mathbb{P}(X_n = x_n \mid X_{n-1} = x_{n-1}, \ldots, X_0 = x_0)$$
$$= \mathbb{P}(X_n = x_n \mid X_{n-1} = x_{n-1}) \quad (2.8),$$

with $\mathbb{P}(X_n | X_{n-1}) = \frac{\mathbb{P}(X_n, \ldots, X_0)}{\mathbb{P}(X_{n-1}, \ldots, X_0)}$ (2.9).

If the Markovian property is satisfied, then if we take as previously the example of a symbol sequence of length $L = 3$, then we must find that:

$p(x_{0,3} = 0 \mid x_{0,2} = 0, x_{0,1} = 0)$
$= p(x_{0,3} = 0 \mid x_{0,2} = 0, x_{0,1} = 1)$
$= p(x_{0,3} = 0 \mid x_{0,2} = 1, x_{0,1} = 1)$
$= p(x_{0,3} = 0 \mid x_{0,2} = 1, x_{0,1} = 0)$

and:
$p(x_{0,3} = 1 \mid x_{0,2} = 0, x_{0,1} = 0)$
$= p(x_{0,3} = 1 \mid x_{0,2} = 0, x_{0,1} = 1)$
$= p(x_{0,3} = 1 \mid x_{0,2} = 1, x_{0,1} = 1)$
$= p(x_{0,3} = 1 \mid x_{0,2} = 1, x_{0,1} = 0)$.

In other words, if the occurrence of symbol sequences of given size $L$ only depends on the present state, not on the sequence of events that preceded it. We decide to investigate these properties and emphasized them in Table2.

|       | $p(0\|0,\star)$ | $p(0\|1,\star)$ | $p(1\|0,\star)$ | $p(1\|1,\star)$ |
|-------|---------|---------|---------|---------|
| .     | 23.0264% | 26.7324% | 26.7324% | 23.5089% |
| 0,0   | 4.7725% | 6.5309% | 5.7409% | 5.9821% |
| 0,1   | 5.7409% | 7.5822% | 6.7721% | 6.6371% |
| 1,0   | 6.5011% | 6.7556% | 7.6120% | 5.8636% |
| 1,1   | 6.0119% | 5.8636% | 6.6073% | 5.0261% |
| 0,0,0,0 | 0.9751% | 1.3436% | 1.1806% | 1.2732% |
| 0,0,0,1 | 1.1806% | 1.6356% | 1.4363% | 1.4884% |
| 0,0,1,0 | 1.3341% | 1.8640% | 1.6054% | 1.6977% |
| 0,0,1,1 | 1.2828% | 1.6877% | 1.5186% | 1.5228% |
| 0,1,0,0 | 1.3720% | 1.8413% | 1.6498% | 1.6677% |
| 0,1,0,1 | 1.5675% | 2.1923% | 1.9118% | 1.9106% |
| 0,1,1,0 | 1.4801% | 1.9149% | 1.7088% | 1.6518% |
| 0,1,1,1 | 1.3213% | 1.6337% | 1.5017% | 1.4070% |
| 1,0,0,0 | 1.3760% | 1.4608% | 1.6033% | 1.3008% |
| 1,0,0,1 | 1.6459% | 1.7229% | 1.9057% | 1.4976% |
| 1,0,1,0 | 1.8280% | 1.9239% | 2.2057% | 1.6544% |
| 1,0,1,1 | 1.6513% | 1.6480% | 1.8972% | 1.4107% |
| 1,1,0,0 | 1.5018% | 1.4746% | 1.6818% | 1.3239% |
| 1,1,0,1 | 1.6871% | 1.6538% | 1.8848% | 1.4114% |
| 1,1,1,0 | 1.5016% | 1.4882% | 1.6268% | 1.2471% |
| 1,1,1,1 | 1.3214% | 1.2471% | 1.4139% | 1.0438% |

**Table2:** Conditional probabilities of symbol sequences with length $L = \{2, 4, 6\}$ on minute time scales.

|       | $p(0\|0,\star)$ | $p(0\|1,\star)$ | $p(1\|0,\star)$ | $p(1\|1,\star)$ |
|-------|---------|---------|---------|---------|
| .     | 19.8812% | 25.7106% | 25.6946% | 28.7137% |
| 0,0   | 3.6787% | 4.9639% | 5.0602% | 6.1687% |
| 0,1   | 5.0602% | 6.9237% | 6.0723% | 7.6466% |
| 1,0   | 5.0763% | 6.8434% | 6.8112% | 6.9719% |
| 1,1   | 6.0562% | 6.9719% | 7.7590% | 7.9357% |
| 0,0,0,0 | 0.6588% | 0.9320% | 0.8677% | 1.2213% |
| 0,0,0,1 | 0.8677% | 1.3338% | 1.2856% | 1.5748% |
| 0,0,1,0 | 0.9963% | 1.1249% | 1.3338% | 1.6069% |
| 0,0,1,1 | 1.1570% | 1.5748% | 1.5587% | 1.7676% |
| 0,1,0,0 | 0.9320% | 1.2373% | 1.2695% | 1.5266% |
| 0,1,0,1 | 1.3980% | 1.9765% | 1.4623% | 2.0730% |
| 0,1,1,0 | 1.3980% | 1.7676% | 1.6552% | 2.0247% |
| 0,1,1,1 | 1.3177% | 1.9444% | 1.6873% | 2.0247% |
| 1,0,0,0 | 0.9963% | 1.3980% | 1.2695% | 1.3980% |
| 1,0,0,1 | 1.2052% | 1.6069% | 1.4945% | 1.7676% |
| 1,0,1,0 | 1.2695% | 1.5909% | 1.9444% | 2.0087% |
| 1,0,1,1 | 1.6069% | 2.2497% | 2.1051% | 1.7998% |
| 1,1,0,0 | 1.2052% | 1.3498% | 1.7998% | 1.8158% |
| 1,1,0,1 | 1.8480% | 1.8801% | 1.9926% | 1.9283% |
| 1,1,1,0 | 1.4302% | 1.7837% | 1.7998% | 1.9605% |
| 1,1,1,1 | 1.5748% | 1.9605% | 2.1694% | 2.2336% |

**Table3:** Conditional probabilities of symbol sequences with length $L = \{2, 4, 6\}$ on 4-hours time scales.

|       | $p(0\|0,\star)$ | $p(0\|1,\star)$ | $p(1\|0,\star)$ | $p(1\|1,\star)$ |
|-------|---------|---------|---------|---------|
| .     | 17.6007% | 26.1821% | 26.0946% | 30.1226% |
| 0,0   | 2.9825% | 4.2982% | 3.9474% | 6.2281% |
| 0,1   | 4.0351% | 8.0702% | 6.5789% | 7.5439% |
| 1,0   | 5.2632% | 5.3509% | 7.1053% | 8.4211% |
| 1,1   | 5.3509% | 8.4211% | 8.5088% | 7.8947% |
| 0,0,0,0 | 0.6151% | 0.8787% | 0.7909% | 0.7030% |
| 0,0,0,1 | 0.7909% | 0.5272% | 0.7909% | 1.8453% |
| 0,0,1,0 | 0.6151% | 1.7575% | 1.3181% | 1.4938% |
| 0,0,1,1 | 0.9666% | 1.1424% | 1.0545% | 2.1968% |
| 0,1,0,0 | 0.8787% | 0.9666% | 0.8787% | 1.5817% |
| 0,1,0,1 | 1.1424% | 2.3726% | 2.3726% | 2.1968% |
| 0,1,1,0 | 0.5272% | 2.1090% | 1.2302% | 1.4938% |
| 0,1,1,1 | 1.4938% | 2.6362% | 2.1090% | 2.1968% |
| 1,0,0,0 | 0.6151% | 0.9666% | 0.7909% | 1.5817% |
| 1,0,0,1 | 1.1424% | 1.6696% | 1.7575% | 2.0211% |
| 1,0,1,0 | 1.1424% | 1.4060% | 2.1968% | 2.3726% |
| 1,0,1,1 | 2.3726% | 1.3181% | 2.3726% | 2.4605% |
| 1,1,0,0 | 0.7909% | 1.5817% | 1.8453% | 2.0211% |
| 1,1,0,1 | 0.9666% | 2.3726% | 1.7575% | 2.4605% |
| 1,1,1,0 | 1.6696% | 2.9877% | 2.2847% | 1.4938% |
| 1,1,1,1 | 1.9332% | 1.4938% | 2.5483% | 1.9332% |

**Table4:** Conditional probabilities of symbol sequences with length $L = \{2, 4, 6\}$ on daily time scales.

In both Table2,3&4 we emphasize the list of conditional probabilities for the respective time scales minutes, 4-hours, and daily with different length of partition starting from 2 to 6 with a step size of two. The idea once again is to investigate the Non-Markovian behavior of the Standard&Poor500 log returns dynamics on different time scales to figure out the presence or not of memory length according to different sizes of symbol sequences $L = \{2, 4, 6\}$. Thus, if we take as an example in Table2 the value in the column $p(0|0,\star)$, this represents the probability that the next event is 0 (i.e. negative returns) knowing that the previous event was also 0 (i.e. also negative returns) and was following by a series of symbol represented by $\star$ that correspond to series of bits $\{0;1\}$ whose the size depends on the corresponding value of $L$.

Thus if ★ corresponds to the row with symbol sequence of {0,0,0,0}, it means that the corresponding size is $L = 6$ and the conditional probability becomes $p(0|0,0,0,0,0)$.

Now, if we analyze the results within Table2,3&4, we assess that on smaller time scales (i.e. minutes) the non-Markovian character for small length $L = 2$ is observable since the perfect equality between respective conditional probabilities is not respected with $p(0|0) \neq p(1|1)$ for example; however the reciprocal is also true with $p(0|1) = p(1|0)$ meaning that we cannot easily conclude that this symbol sequence is governed by a purely non-Markovian process. We also observe a certain dispersion of 1.7% between conditional probabilities ranging from 23% to 26.7%.

Consequently, on small time scales, Standard&Poor500 log returns dynamics does depend on present event for some future events and exhibit partially some memory length and predictability.

On larger time scales (i.e. 4-hours and daily) the assessment is more straightforward. Indeed, for the same symbol sequence length $L = 2$, we observe that for both time scales we have $p(0|0) \neq p(1|1) \neq p(1|0) \neq p(0|1)$ meaning that on larger time scales the Standard&Poor500 log returns dynamics also exhibit non-Markovian character with dependence to previous events. Indeed, if the $p(1|0)$ is slightly different to $p(0|1)$ the dispersion of probabilities is enough significant (3.2% for the 4hours symbol sequence and 4.57% for the daily symbol sequence) to deduct unequal conditional probabilities.

Moreover, we assess that the variability of conditional probabilities increases as the time scales become larger, and we also assess that the highest conditional probability for both time scales is $p(1|1)$. Namely, the probability that the next event is a positive return when the previous event was also a positive return is the most likely, which is consistent with the long-term bullish growth of the Standard&Poor500 over the studied period of time.

For other length, namely $L = \{4, 6, 8\}$ we propose to analyze the results through a whisker plot in order to deal with more convenient descriptive statistics than table2,3&4 due to the larger size of symbol sequences. We recall that for $L = 8$, we have $\Phi(L = 8) = 2^8 = 256$ possible symbol sequences.

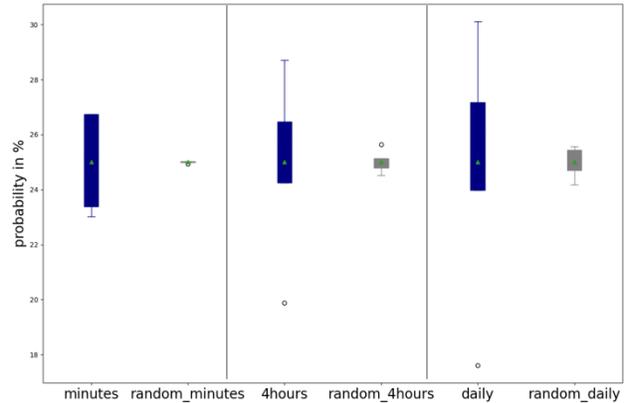

**Figure7:** In blue, Box-Plot Graph representing the statistical features of symbol sequences of length L = 2 on different timescales (minutes, 4hours, daily) of the Standard&Poor500's log returns dynamics. In grey, Box-Plot Graph representing the statistical features of symbol sequences of length **L = 2** on different time-scales (minutes, 4hours, daily) of perfectly random dynamics symbols.

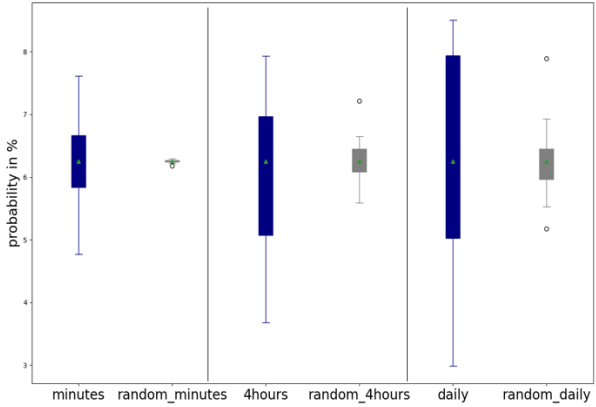

**Figure8:** In blue, Box-Plot Graph representing the statistical features of symbol sequences of length **L = 4** on different timescales (minutes, 4hours, daily) of the Standard&Poor500's log returns dynamics. In grey, Box-Plot Graph representing the statistical features of symbol sequences of length **L = 4** on different timescales (minutes, 4hours, daily) of perfectly random dynamics symbols.

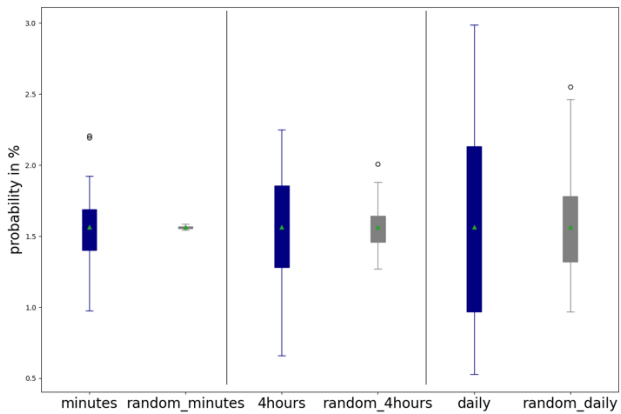

**Figure9:** In blue, Box-Plot Graph representing the statistical features of symbol sequences of length **L = 6** on different timescales (minutes, 4hours, daily) of the Standard&Poor500's log returns dynamics. In grey, Box-Plot Graph representing the statistical features of symbol sequences of length **L = 6** on different timescales (minutes, 4hours, daily) of perfectly random dynamics symbols.

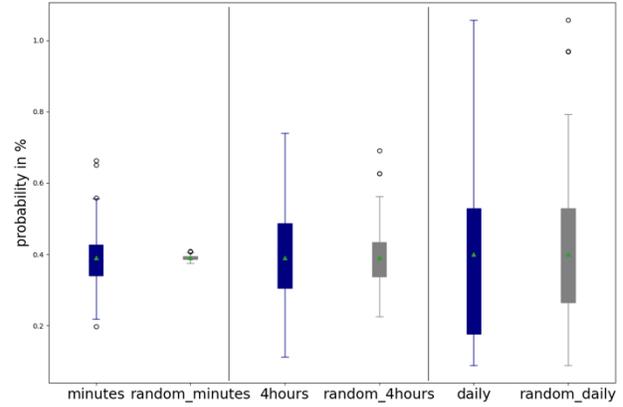

**Figure10:** In blue, Box-Plot Graph representing the statistical features of symbol sequences of length **L = 8** on different timescales (minutes, 4hours, daily) of the Standard&Poor500's log returns dynamics. In grey, Box-Plot Graph representing the statistical features of symbol sequences of length **L = 8** on different timescales (minutes, 4hours, daily) of perfectly random dynamics symbols.

Figure7,8,9&10 emphasized a series of whisker plot for different respective symbol sequences length $L = \{2, 4, 6, 8\}$. In blue we represent the statistical features of symbol sequence from the Standard&Poor500 log returns dynamics, and we compare them with the statistics coming from symbol sequences of same length but purely random.

When $L = 2$, we enjoy more clear statistics than from Table2,3&4. On smaller time scales (minutes), conditional probabilities are much more disperse on the Standard&Poor500 symbol dynamics than random symbol dynamics with larger interquartile size. This is emphasized within Table5 that represent the ratio between the interquartile length of the Standard&Poor500 dynamics and the length of random dynamics. We assess that on minutes time scales and for $L = 2$, the length is 122 times larger emphasizing a strong dispersion within probabilities.

| Ratio ∅ | L = 2 | L = 4 | L = 6 | L = 8 |
|---|---|---|---|---|
| Minutes | 122.7 | 36 | 20.12 | 10.96 |
| 4hours | 6.51 | 5.14 | 3.11 | 1.88 |
| Daily | 4.29 | 6.05 | 2.52 | 1.33 |

**Table5:** Comparison for different length $L$ and on different time scales of the interquartile (IQR) length ratio that we call $\emptyset = \frac{length\ of\ IQR\ from\ SP500}{length\ IQR\ from\ random\ series}$.

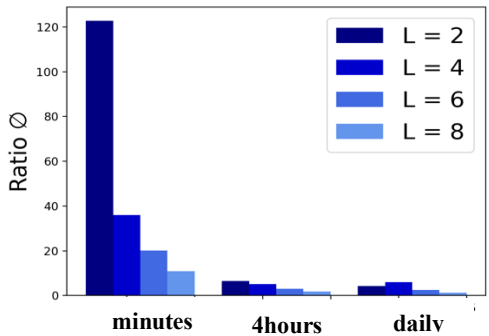

**Figure11:** Bar chart of interquartile (IQR) length ratio that we call $\emptyset = \frac{length\ of\ IQR\ from\ SP500}{length\ IQR\ from\ random\ series}$.

When $L = \{4, 6, 8\}$, we still observe for both the 4hours and daily time scales a slight difference between their respective median the one from random symbol sequence. Moreover, on every time scales, it is clear that the strong dispersion of conditional probabilities from Stadard&Poor500 dynamics persists, and the length of their respective interquartile is much longer than the one from random symbol sequence.

However, this difference in interquartile length decreases as the length $L$ of the symbol sequence increases. This is emphasized in both Table5 and Figure 11, on every time scale, as much as we increase the number of past events, the Standard&Poor500 tends to converge faster towards similar statistics than purely random, although for $L = 8$, the dynamics still exhibit strong non-Markovian behaviors. But it remains important to understand that the dependence of a future event on a series of past events decreases as the length of historical data increases. And this decrease is faster on smaller time scales than on longer ones (Table5).

In short, we understand from previous analysis that more the length of historical data increase, less the next event become dependent on it and less it becomes predictable. In other words, the predictability of log returns dynamics depends on:
- the time scales: larger the time scales, the more the dynamics exhibit non-Markovian features.
- the length of historical events under consideration: smaller the length, the stronger the dependence with the next event.

## 4. Entropy and Complexity

Measuring the memory length or the degree of predictability of future log returns conditioned to past log returns as we did in section 3 is actually equivalent to trying to measure our degree of uncertainty on predicted future log-returns based on different length of past data. More the conditional probabilities of future events i.e. {0,1} conditioned to a series of past events for a given length are varying from each other (strong dispersion), lower is our uncertainty about what will happen next.

This concept of considering the degree of predictability as being associated with the degree of uncertainty is actually the fundamental premise of Shannon works on Information theory. Shannon defined Information as being closely associated with uncertainty:

$$I(r^{(L)}) = \sum_i p_i^{(L)} \ln p_i^{(L)} \quad (4.1),$$

with $I$ the Shannon Information or the function of distribution of every possible event $i$, $r^{(L)} = \Phi(L)^{-1} = 2^{-L}$ the scaling size of the series and $p_i$ the probability of this event. Since $0 \leq p_i \leq 1$ then $I(r^{(L)}) \leq 0$ and it reaches its maximum 0 when the knowledge about a future event is maximum. In other hands $I(r^{(L)})$ reaches its minimum $-\ln\Phi(L) = \ln(r^{(L)})$ when all events $i = 1, 2, \ldots, \Phi(L)$ are distributed with a uniform distribution, namely, $p_i = \frac{1}{\Phi(L)}$ (4.2), meaning that since all events have the same probability to occur, we have a perfect uncertainty on what will happen in future.

We are more interested in the negative Shannon information measure also called entropy such that:

$$S(r^{(L)}) = -I(r^{(L)}) \quad (4.3).$$

Here, the Shannon entropy is a measure of surprise about an event, meaning our degree of uncertainty. In other words, a low entropy sample exhibits patterns, ordered and complex structures being more or less predictable whereas high entropy sample will tend to be uniformly distributed and perfectly random. Therefore, we can also measure the predictability or the memory length of the Standard&Poor500 by measuring its entropy.

To this end we propose to apply to the Generalized Information also called the Rényi Information [18] so-called because it allows generalizing the Shannon information by taking into account an arbitrary and non-uniform number called $q$ that determines the weight or importance assigned to the probability of events; the purpose being to characterize the degree of complexity of a sample by computing a full spectrum of generalized Information or Entropy.
We define the Rényi generalized Information by:

$$I(r^{(L)})_q = \frac{1}{q-1} \log \sum_i^{\Phi(L)} \left(p_i^{(L)}\right)^q \quad (4.4).$$

As mentioned above, we can regard the Rényi information as a generalization of the Shannon information for $q \to 1$ as following:

$$\lim_{q \to 1} I(r^{(L)})_q = \sum_i^{\Phi(L)} \sum_i p_i^{(L)} \ln p_i^{(L)} \quad (4.5)$$

$$\lim_{q \to 1} I(r^{(L)}) = I(r^{(L)}) \quad (4.6)$$

Thus, in order to measure the level of complexity of the Stadard&Poor500, we propose to analyze first how the Rényi information defined by equation (4.4) evolves and grows on different time scales according to the inverse change in length of the symbol sequences that we have defined by $r^{(L)}$, and second, we want to measure the changing of the negative Rényi information on different time scales when the length $L$ grows.
To that end, we will define two other important concepts.

The first one is the Rényi dimension that measures the change in Rényi information defined by the following relation:

$$D_q = \lim_{r^{(L)} \to 0} \frac{I(r^{(L)})_q}{\ln r^{(L)}} \quad (4.6),$$

$$D_q = \lim_{r^{(L)} \to 0} \frac{1}{\ln r^{(L)}} \frac{1}{q-1} \log \sum_i^{\Phi(L)} \left(p_i^{(L)}\right)^q \quad (4.7)$$

The Rényi dimension is a decreasing function of $q$ and will remain finite as long as $r^{(L)} \to 0$. Depending on the value of $q$, the Rényi dimension reveals different characteristics of the symbol sequence, such as its degree of fractality or multifractality; for example, for $q = 0$ we find back the capacity dimension, $q = 1$ we find back the Shannon dimension and for $q = 2$ we find back the Hausdorff dimension, i.e. they are all special cases of the Rényi dimension.

The second one is the Rényi entropy that corresponds to the negative Rényi information defined by the following relation:

$$S_q = \lim_{L \to \infty} \frac{-I(r^{(L)})_q}{L} \quad (4.8)$$

$$S_q = \lim_{L \to \infty} \frac{1}{1-q} * \frac{1}{L} * \log \sum_i^{\Phi(L)} \left(p_i^{(L)}\right)^q \quad (4.9)$$

The Rényi entropies have the same dependence on $q$ as $D_q$. Moreover, in the same way than the Rényi dimension, we find back some special entropies such as the Kolomogrov-Sinai entropy when $q \to 1$ such that $S_1 = \lim_{L \to \infty} \frac{-I(r^{(L)})}{L}$ (4.10) and for $q = 0$ we get the topological entropy entropy such that $S_0 = \lim_{L \to \infty} \ln \frac{\Phi(L)}{L} = \lim_{L \to \infty} \ln 2^L = \ln 2$ (4.11).

We propose to highlight different Rényi entropies for different values of $q$ ranging from -40 to 40, on different time scales for different length $L$ and according to the equation (4.9).

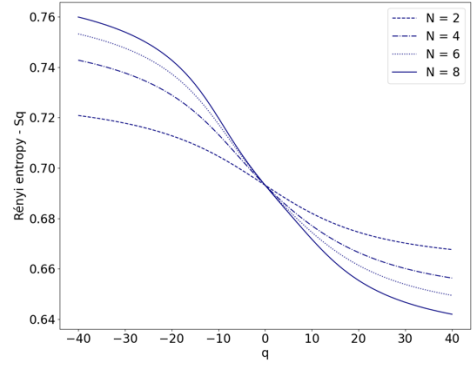

**Figure12:** Rényi entropy of Standard&Poor500 symbol sequence dynamics for $L = \{2, 4, 6, 8\}$ on minutes tine scales.

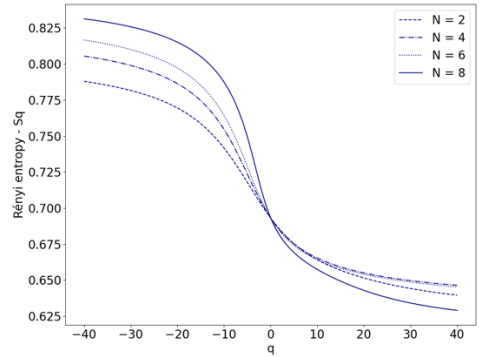

**Figure13:** Rényi entropy of Standard&Poor500 symbol sequence dynamics for $L = \{2, 4, 6, 8\}$ on 4hours tine scales.

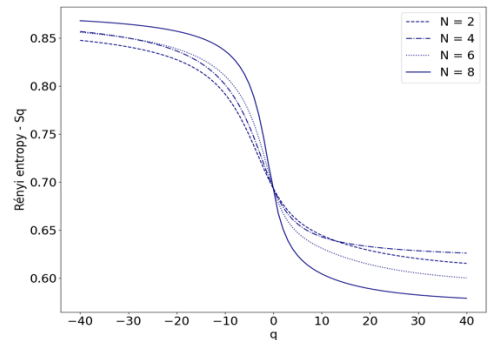

**Figure14:** Rényi entropy of Standard&Poor500 symbol sequence dynamics for $L = \{2, 4, 6, 8\}$ on daily tine scales.

Figure12,13&14 emphasized the negative Rényi information and its dependence on $q$. We observe that for every different time scale, we have a strong dependence of entropies on $q$ with relatively sharp decreases as $q$ increases. Moreover, the larger the length of the symbol sequence, the stronger is the dependence, meaning that the complexity of the Standard&Poor500 increases as the number of historical events taken into consideration increases.

We also observe that Rényi entropies have a stronger dependence on $q$ on larger time scales than smaller time scales. This observation is highlighted by Figure14, for both $L = 2$ and 8, the Rényi entropy is much flatter on minutes time scales than on daily time scales resulting in a stronger conditional probabilities dependency on $q$.

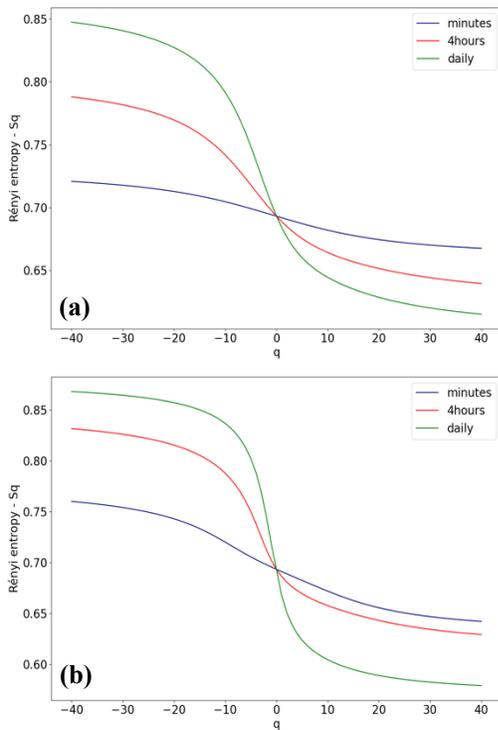

**Figure15:** Comparison **of** Rényi entropies of Standard&Poor500 symbol sequence dynamics for $L = 2$ (a) and for $L = 8$ (b) on different tine scales.

We also observe from Figure15 that for negative values of $q$, the Renyi entropies on minute time scales are lower than on larger time, meaning that symbol sequence on minute time scales exhibit less uncertainty for negative values of $q$ than on larger time scales.

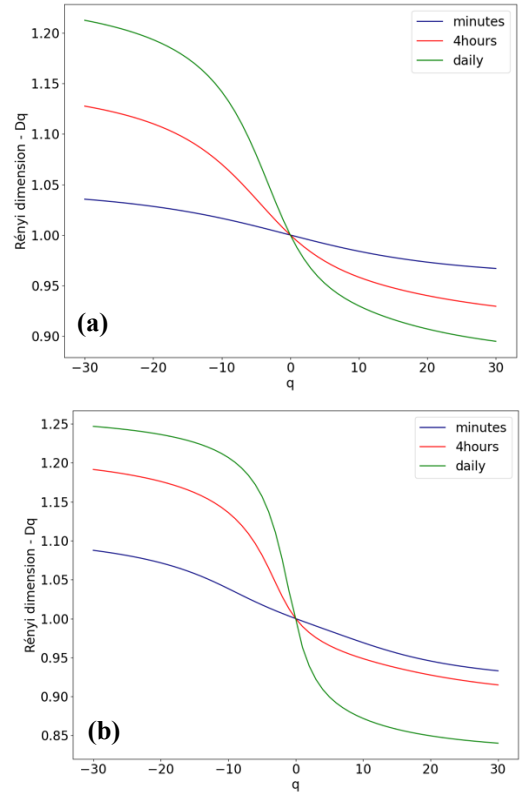

**Figure16:** Comparison of Rényi dimension of Standard&Poor500 symbol sequence dynamics for $L = 2$ (a) and for $L = 8$ (b) on different tine scales.

Figure16 represents the Rényi dimension in accordance with relation (4.7). We can also observe first that the larger the length $L$, the larger the dependence of the dimension on different values of $q$ on every time scales, second, the larger the time scales, the stronger the dependence on $q$. We also note that on minutes time scales and for small length $L$, the Rényi dimension is relatively flat and close to a dimension of 1, emphasizing fewer complex structures and highlight more mono-fractal property. Consequently, we can deduct that the complexity of the Standard&Poor500 and its multifractality property is increasing as the size of the time scales and as the size of the symbol sequences are both increasing.

We propose to go a step further and provide a new way of interpreting the generalized by dimension by introducing the Spectrum of Scaling Indices $f(\alpha)$ introduced by Halsey [13] in 1986. The latter in his article introduce a change of variables through a Legendre transformation [13] $(q, \tau(q))$ with $\tau(q) = D_q(q-1))$ to new variables $(\alpha, f)$ which enable us to have a new interpretation of the generalized Rényi dimension, through the singularity dimension $\alpha$ such that for $q \neq 1$ we have [13, 7, 15]:

$$\begin{cases} D_q = \dfrac{f(\alpha) - q\alpha}{1-q} & (4.12) \\ f(\alpha) = D_q(1-q) + q\alpha & (4.13) \end{cases}$$

As explained by H.Salat, R. Murcio and E. Arcaute [8] if the Generalized dimension only gives global measurements of the whole data, in contrast, the multifractal spectrum gives one dimension for each set where the data scales similarly. In a sense, the variable $q$ selects different resolutions with higher values of $q$ selecting a local scaling $\alpha(q)$ of lower order. The variable $f(\alpha(q))$ then gives the local fractal dimension at resolution $q$. And it is trivial to see that the spectrum's peak is achieved for $q = 0$, where $f(\alpha_0) = D_0$ namely the fractal dimension of capacity.

To compute $f(\alpha(q))$, we propose to deal with the Chhabra's method [1] that provide a recipe to avoid the Legendre transform of $\tau(q)$ such that we obtain the following relations:

$$f(q) = \lim_{r \to 0} \frac{\sum \mu_i^q(r) \log(\mu_i^q(r))}{\log(r)} \quad (4.14),$$

$$\alpha(q) = \lim_{r \to 0} \frac{\sum \mu_i^q(r) \log p_i(r)}{\log(r)} \quad (4.15).$$

Chhabra considers a subset, a grid made up of several cells of unit size $r$ covering a domain $D \subset \mathbb{R}^n$ on which is defined a fractal characterized by a finite- multifractal $\mu$. This multi-fractal measurement $\mu$ is characterized by a distribution such that around any point $x \in D$, the measurement of a cell of the grid of length $r$ around $x$ is proportional. To the power law $r^\alpha$, with $\alpha$ corresponding to the scaling exponent and can take any given values, under the condition that $r$ being small enough that the sets formed by the cells of the grid are defined by the same scale $\alpha$.

More generally, the multi-fractal measure $\mu$ for sufficient small values of $r$ is defined by: $\mu_r(x) \sim r^{\alpha_x}$ (4.16) for a scale factor $\alpha_x$ with $x \in D$ and $\mu_r(x)$ being the measure of a cell of length $r$. Chhabra also interprets the multi-fractal measure $\mu_r(x)$ in a probabilistic way, he calls $N$ the number of cells from the grid necessary to cover the subset $D$, and gives the probability of observing any cell as $p_i \coloneqq \dfrac{N_i}{N} = \int_{i\text{ème cases}} d\mu(x)$ (4.17) with $p_i = r^{\alpha_i}$ (4.18) the probability that a part of the fractal is measured by an *ith* cell. Since $\mu$ is the multifractal measure of the subset D, Chhabra compute the local multi-fractal measure associated to each cell $i$ as: $\mu_i(r) =$

$\frac{p_i}{\Sigma_j p_j}$ (4.19), and then propose the following relation:

$$\mu_i^q(r) = \frac{p_i^q}{\Sigma_j p_j^q} \text{ (4.20).}$$

Finally, the Legendre transform can be directly integrated in the calculation of $f$ and $\alpha$ to obtain equation (4.14) and (4.15).

We also mention that the spread of $\alpha$ indicates the variety of scaling present in the sample while the value of $f(\alpha)$ indicates the strength of the contribution of each $\alpha$.

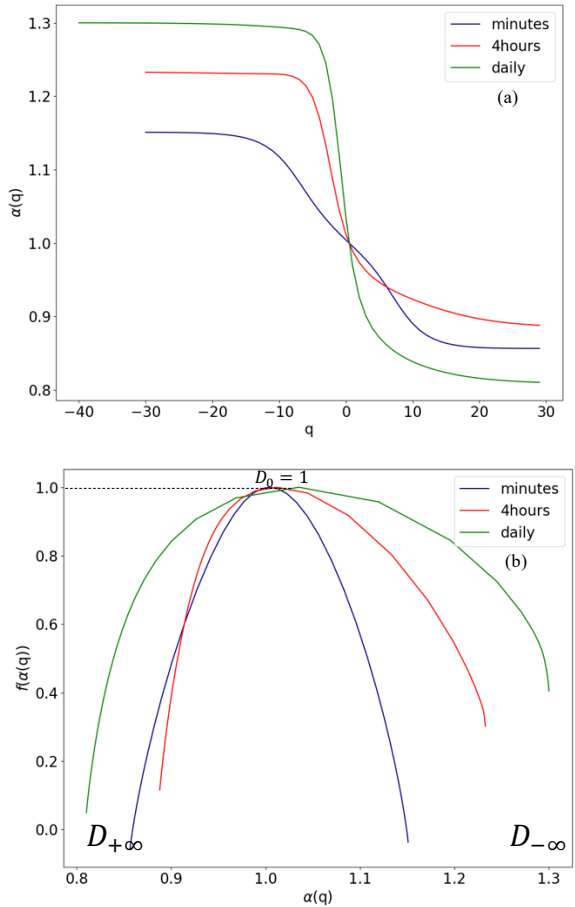

**Figure17:** (a) Comparison of curves $\alpha(q)$ versus $q$ defined by equation (4.15) on different time scales and (b) comparison of functions $f(\alpha(q))$ against $\alpha(q)$ defined by equations (4.14) for scales values ranging from 2 to 9 and $q$ ranging from -30 to 30 on different time scales.

Figure17 represents a clear picture of the structural differences of the Standard&Poor500 dynamics on different time scales. The strong difference in value and shape between minutes and daily time scales is an evidence of the Standard&Poor500's multifractal nature with no homonymity of behaviors among time scales.

Firstly, the singularity exponent $\alpha(q)$ accounts for the balance between symbol dynamics with smaller and larger lengths. We observe that for positive $q$ values, the $\alpha(q)$ are relatively lower and the difference between time scales get smaller, meaning that the number of symbol sequences with strong probabilities are not that common. For negative values of $q$, we observe larger values of $\alpha(q)$ with an increase in difference between different time scales, confirming the fact that there is no major symbol sequence pattern that dominates among time scales.

Second, the multifractal spectrum curves (Figure16. (b)) represent the distribution of symbol sequences for different scaling factors. The right part of each curve (related to negative $q$) becomes more and more widespread as we move forward on time scales with $f\alpha(q)$ taking greater values as the time scales increase. The left part of each curve (related to positive $q$) is less widespread especially regarding the 4-hours and minutes time scales but remain clear when comparing with the daily time scales. This is a clear indicator of how the Standard&Poor500 dynamics are becoming more different through different time scales, exhibiting different statistics, prove its multifractal specificity.

Finally, we almost observe a perfect symmetry of $f(\alpha(q))$ on minutes time scales, meaning that for $q \in [-\infty; +\infty]$ the symbol sequences are quite self-similar with typically less varieties of conditional probabilities between symbol sequences, which confirm our analysis in section 3&4. However, concerning the 4-hours and daily time scales, we observe a strong concentration of $f(\alpha(q))$ when $q \to -\infty$, meaning that there are greater varieties of symbol sequences with low than high probabilities, also confirming our previous assessment.

## 5. Conclusion

Standard&Poor500's predictability and complexity characteristics have been examined on different time scales in a coarse-grained way using the symbolic dynamics method and under the prism of Information theory with the concept of entropy and uncertainty.

We first gave evidence of the non-Markovian behaviors of the Standard&Poor500 dynamics by highlighting unequal conditional probabilities among symbol sequences of the same length, and this on different time scales. Proving the presence of memory process and dependency characteristics conditioned to historical log-returns. We also found that if on larger time scales, positive log-returns were more likely to be followed by positive log-returns, on small time scales (minutes), positive (negative) log-returns were more likely to be followed by negative (positive) log-returns.

Second, from these conditional probabilities of the corresponding symbol sequences calculated on different timescales, a non-trivial Rényi entropy and Dimension spectrum have been found, with stronger dependency on values of $q$ as the time scales became larger, with a sharp decrease of the Dimension $D_q$ as $q$ increases, and a relatively flat Dimension on small time scales (minutes).

Third, we computed the Spectrum of Singularity Indices. Similar evidence was found with more homogeneity and symmetry of $f(\alpha(q))$ on small time scales (minutes) and more varieties of $f(\alpha(q))$ values on larger time scales according to different values of $q$ with strong concentration of $f(\alpha(q))$ when $q \to -\infty$.

We conclude that the Standard&Poor500 dynamics are becoming more different through different time scales, exhibiting different statistics, more dependency, and more complexity as the time scales become larger.